\documentclass[aip,graphicx]{revtex4-1}
\usepackage{graphicx}% Include figure files
\draft % marks overfull lines with a black rule on the right

\begin{document}
%\preprint{AIP/123-QED}

\title{Propagation characteristics of complementary split-ring resonators excited by internal  cylindrical wave}% Force line breaks with \\
%\thanks{This work was supported by the Fundamental Research Funds for the Central Universities (K50510040015).}
\author{ZHu Yanwu }
\email{zhuyanwu@xidian.edu.cn.}
 \author{Lu Shan }%
\author{Xu SHejiao}%
%\author{Tian Jin}%
\author{Xiang Min}%
%\author{Qiu Yang}%
\affiliation{
School of Mechano-Electric Engineering, Xidian University,\\ Xi'an 710071, China}

\date{\today}% It is always \today, today,

\begin{abstract}
For a complementary split-ring resonators (CSRRs) etched on one screen of two infinite and perfectly conducting plates under cylindrical plane wave illumination, the transmission  coefficients analysis were performed through finite-difference time-domain simulations. For the single slot ring CSRR, two transmission dips are observed for the direction parallel to aperture at resonance whereas two enhanced transmission for the direction normal to the aperture. For double slot ring CSRR, the transmission coefficients appear more anisotropic. The results provided considerable insight into the electromagnetic response of CSRR and would be very helpful for developing new electronic devices such as filters.
\end{abstract}

\maketitle

%\section{\label{sec:level1}First-level heading:\protect\\ The line
%break was forced \lowercase{via} \textbackslash\textbackslash}
\section{Introduction}
Electromagnetic metamaterials have attracted enormous research interest since the initial  experiment demonstrations of their unique properties reported by Smith \emph{et al.}.\cite{Smith} Various metamaterials have been proposed to realize their exotic properties, such as negative permittivity, negative permeability and double-negative, from the microwave region to the optical region. Negative values of the permittivity are attainable through thin-wire lattices\cite{Pendry96} and negative permeability through such as split-ring resonators (SRRs)\cite{Pendry99} or cut-wire pairs.\cite{Powell08} Recently, based on the Babinet principle and the duality concept, it was demonstrated by F. Falcone that the complementary split-ring resonators (CSRRs) excited by axial electric field could exhibit negative permittivity upon their resonance, which behave as an electrical dipoles.\cite{Falconeprl} Since their introduction, CSRRs have undergone many studies to accommodate particular applications. A super-compact stopband microstrip structure is proposed by an array of CSRRs etched on the ground plane, underneath the conductor strip.\cite{Falcone04} Planar electric split ring resonator (eSRR) metamaterials and their corresponding inverse structures are designed and characterized computationally and experimentally utilizing finite element modeling and THz time domain spectroscopy\cite{chen}. A two-dimensional 2D planar gradient index circuits formed by CSRR has been shown to be useful in the waveguide environment.\cite{R} The magnetic counterpart of the CSRR, the complementary electric LC  resonator, achieves a purely magnetic response with no cross coupling.\cite{Thomas} An ultrathin chiral metamaterial slab stacked with twisted CSRRs was proposed for highly efficient broadband polarization transformation.\cite{wei} The CSRRs as an electrically small resonator have been found to be very appropriate for the filters design or frequency selective surface.\cite{Izquierdo,Francisco} More recently, for CSRR is a natural filter for electric fields polarized normal to the slot ring, Omar M. Ramahi \emph{et al.}. presented a novel technique for switching noise mitigation in printed circuit boards (PCBs) which is achieved through etching concentrically cascading CSRRs on only a single metallic layer of the PCB.\cite{Omar11} It is shown here that  an ultrawideband suppression of switching noise from sub-GHz to 12 GHz is achieved.

However, it seemed that the condition when the CSRR illuminated by a cylindrical wave generated by the source located in the origin of ring has not been studied carefully. On one hand, in the previous research on the anisotropic characteristics, the SRR and CSRR is driven by the external illumination which the plane waves are incident in the k direction and two different polarizations are considered: parallel polarization, where the magnetic field H is parallel to the SRR z axis and perpendicular polarization, where it is normal to that axis.\cite{David} On the other hand, the negative effective permittivity related to the CSRR has been discussed in the case that CSRR etched in the ground plane of a microstrip line can be excited by means of an axial time varying electric field.\cite{Falconeprl} Therefore, in this paper, we study the resonance phenomena of CSRR excited by a wire current source located the origin of slit rings, between two plate located in z-axis, which created a cylindric wave with the propagation vector is along the radial direction and a horizontal magnetic field along the xy-plate, by means of numerical modeling using the finite-difference time-domain (FDTD) method.\cite{Elena}
\section{Model and method}
\begin{figure}
\centering
\includegraphics[scale=1,angle=0]{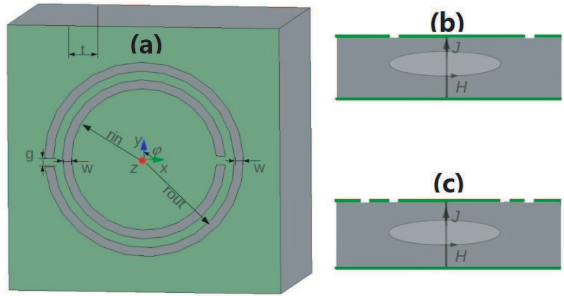}% Here is how to import EPS art
\caption{\label{fig:1}  (a) 3D schematic of simulated model. The parameters of the CSRR are rin=4mm (inner slot ring radius), rout=4.4mm (outer slot ring radius) and w=0.2mm (line width), g=0.2mm (gap width) and t=1.54mm (thickness). (b) XZ cross sections of left model. (c) YZ cross sections of left model. Green indicates
metallic layers.}
\end{figure}

 The model we studied is shown in Fig.~\ref{fig:1}(a). The two infinite and perfectly conducting plates lie in the x-y plane and are located at z = 0 and z =1.54mm. The CSRR is the complementary counterpart of the SRR, and hence it consists on a pair of slot rings with apertures in opposite orientation, etched only on the metallic layer at z=1.54mm. The space between two plates was filled of  the dielectric media with a relative permittivity of 3.4 and lossless. Fig.~\ref{fig:1}(b) and (c) present XZ and YZ cross sections of model, respectively. The FDTD grid spacing is 0.05mm,  which implies very high resolution. Perfectly matched layers were used to terminate boundaries of the computational domain.
  %For simplicity, we assume that the current distribution of the wire source is given by
% %\begin{equation}
%%\left\{
%%J_z(t)=J_0e^{-\alpha(t-t_s)^2}
%%\right\}.
%%\end{equation}.
%\begin{equation*}
%J_z(t)=J_0e^{-\alpha(t-t_s)^2}%%%% ¹«Ê½ÄÚÈÝ
%\end{equation*}
\section{Results and discuss}
\begin{figure}
%\includegraphics[scale=0.2]{Fig2(a)}% Here is how to import EPS art
%\includegraphics[scale=0.2]{Fig2(b)}% Here is how to import EPS art
%\\
%\includegraphics[scale=0.2]{Fig2(c)}% Here is how to import EPS art
%\includegraphics[scale=0.2]{Fig2(d)}% Here is how to import EPS art
\includegraphics[scale=0.7]{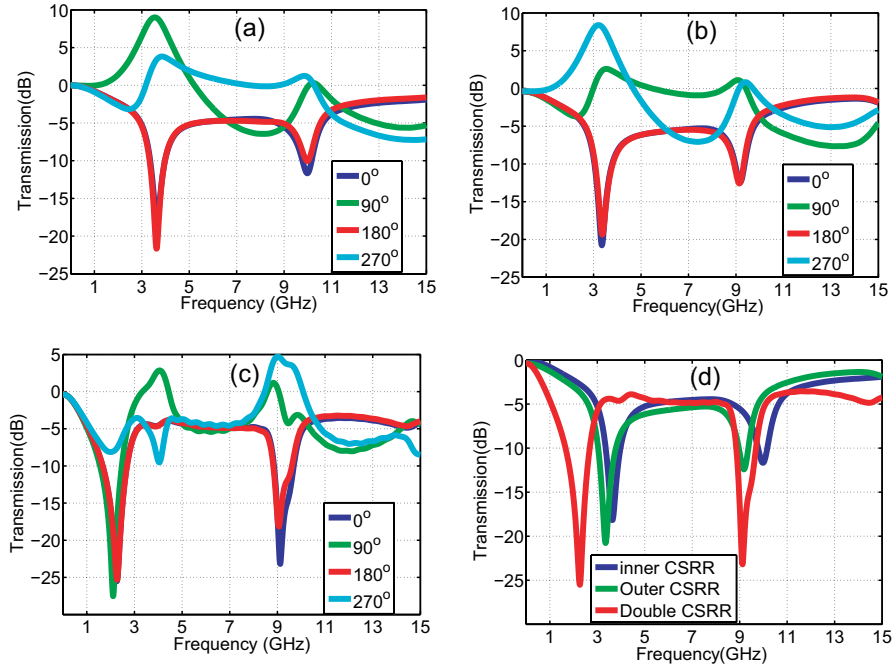}% Here is how to import EPS art
\caption{\label{fig:2} (Color online) Simulate transmission coefficient for different ring.
(a) Inner slot ring over four angles. (b) Outer slot ring over four angles. (c) double slot ring over four angles
(d) different slot ring for  $\varphi$ = 90$^\circ$.}
\end{figure}

Kafesaki \emph{et al.} \cite{Kafesaki} have pointed out that in order to obtain a magnetic resonance which possesses a region with  $\mu<$  0 it is not necessary to have a double-ring SRR, as was probably thought and mostly done both in the experiments and in the simulations. A single ring with a cut (a single-ring SRR) behaves also as a magnetic resonator. In order to examine the  relationship between single slot ring and double slot ring, the inner slot ring, outer slot ring and double slot ring CSRRs were simulated separately in our study. We simulated voltage between the metal sheets of the parallel-plate at the positions around the sample. The fast Fourier transformation of the time-domain data for the voltage was used to obtain the spectrum characterizing EM response of the CSRR.  The spectra was sampled at the distance 8.0mm around source at  $\varphi =$ 0$^\circ$, 90$^\circ$, 180$^\circ$ and 270$^\circ$. Transmission coefficient versus frequency at these spots for three kinds of rings were computed and showed in Fig.~\ref{fig:2}. For the inner slot ring, the first dip which drops than 20 dB was observed at frequency of approximately 3.54 GHz and second dip at frequency of approximately 10 GHz for $\varphi =$90$^\circ$ and 270$^\circ$ in Fig.~\ref{fig:2}(a).  For the outer slot ring, the first dip which drops than 20 dB was observed at frequency of approximately 3.25 GHz and second dip at frequency of approximately 9 GHz for $\varphi =$90$^\circ$ and 270$^\circ$ in Fig.~\ref{fig:2}(b). The cylindric wave ensures that the magnetic fields are configured to be completely in xy plane, which indicates that there is no component of the magnetic field capable of causing a magnetic response by driving circulating currents. Thus, it is associated with a negative  $\epsilon$ regime which correspond to the electric resonance frequency. The LC resonance of an CSRR, $\omega=1/\sqrt{LC}$, is due to the capacitance provided by its circular air loop driven by the incident field acting on the asymmetric structure of the CSRR. Since the capacitance is proportional to the ring radius, smaller radius means smaller capacitance and thus larger resonance frequency. What we are more interesting is, compared with the two narrower and deeper dips in the transmission coefficient for $\varphi =$90$^\circ$ and 270$^\circ$, two broader and weaker peaks appeared at the same frequency for inner slot ring at $\varphi =$0$^\circ$ and for outer slot ring at 180$^\circ$ which corresponded to the aperture direction.  The reason for this enhanced transmission would be explained after the discussion of electric field distributions.

Since a single slot ring acts also as a electric resonator, providing a negative $\epsilon$ regime, what are the benefits from the addition of a second slot ring? In Fig.~\ref{fig:2}(c) we present the transmission through the double slot ring CSRR at four different angles. It can be seen that the transmission of the double slot ring has a drop in the first dip at 1.95GHz on four directions.  Hence, one advantage of the double slot ring CSRR, compared with its single slot ring resonator is making the CSRRs appear more isotropic to the electromagnetic excitation at the lower resonance frequency.  But for the existing of aperture of outer slot ring and the coupling of inner slot ring, the enhanced transmission only occur in the direction for $\varphi =$0$^\circ$  and a dip transmission in the direction for $\varphi =$0$^\circ$ at the  first frequency of outer slot ring. Another advantage of double slot ring CSRR is that the first resonance frequency of the double ring occurs at a relatively lower frequency than the first resonance frequency of the single slot ring (see the first dip in Fig.~\ref{fig:2}(d)).

\begin{figure}
\includegraphics[scale=0.6]{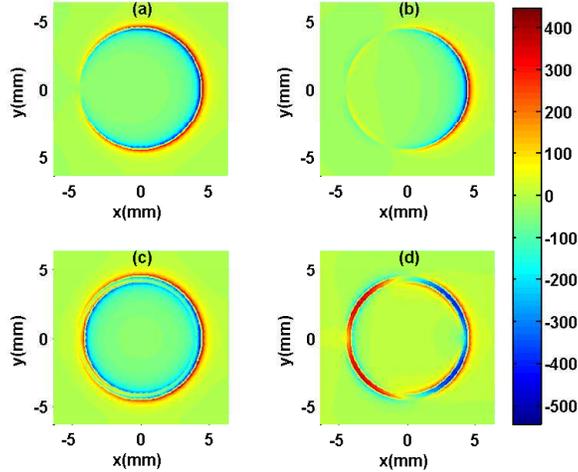}% Here is how to import EPS art
\caption{\label{fig:3}  Electric field component perpendicular to the CSRR plane is
shown in the CSRR plane. Red color represents positive, blue negative and green zero intensity field. (a) First resonant frequency of outer slot ring. (b) Second resonant frequency of outer slot ring. (c) first resonant frequency of double slot ring. (d) Second resonant frequency of double slot ring.}
\end{figure}

In order to understand the nature of resonances, we simulate the amplitude  distributions for oscillations of electric fields in our sample at different resonance frequencies in Fig.~\ref{fig:2}. These distributions were obtained from the xy-plane at z=1.54mm. For lower resonance frequency of single slot ring CSRR, the strongest resonant electric-field are observed almost all around the slot ring edges opposite to the aperture in Fig.~\ref{fig:3}(a). While for the higher resonance frequency of single CSRR the strongest resonant electric-field opposite in phase are observed only around half of the slot ring edges in Fig.~\ref{fig:3}(b). Thus the capacitor formed by oscillations electric field of lower resonate frequency is larger than the capacitor of higher resonance frequency. For lower resonance frequency of double slot ring CSRR, the strongest resonant electric field around the outer slot ring edge and inner slot ring are opposite in direction and same in phase in Fig.~\ref{fig:3}(c). Thus two capacitor are formed independently and the net capacitance is that of the capacitances of each region in series and thus lowers the resonance frequency, corresponding to the lower resonance frequency. For higher resonance frequency of double slot ring CSRR the resonant electric field around the outer and inner slot ring are all divided into two halves and opposite in phase in Fig.~\ref{fig:3}(d). Hence two resonators are formed independently and then grouped to a system. Thus the second dip of the double-slot ring case corresponds essentially to the high frequency resonance of the outer slot ring  and the resonance intensity is approximately twice of the outer slot ring as seen in Fig.~\ref{fig:2}(d). These analysis lead us to suggest that the oscillations of electric field around the slot behavior like a perfect electric-conductor at the resonant frequency and reflect  internal fields between the symmetrized structure resulted the transmission drop and radiated from the aperture direction resulted the enhanced transmission.
\section{Conclusion}
To conclude, we have studied the transmission properties of single slot ring CSRR and double slot ring CSRR unit cells excited by a wire current source. We find except two dips in transmission spectra in the direction parallel to aperture, there are two enhance transmissions in the direction perpendicular to aperture. For the double slot ring CSRR case, the transmission around the CSRR appears more homogeneous to the electromagnetic excitation and the lower resonate frequency decreased than the single slot ring slot. We also analyzed the field distribution in order to observe how the electric field oscillating at the slot ring. By making a comparison between single and double slot ring CSRR the electric resonances can be found easily. This detailed numerical results will be helpful in the theoretical understanding, analysis, development on CSRRs, and possibly help in the investigation of the feasibility of their use in different applications.

This work was supported by the Fundamental Research Funds for the Central Universities (K50510040015).
\nocite{*}
%\bibliography{aipsamp}% Produces the bibliography via BibTeX.
%

\end{document}